**Growth of Knowledge and Entropy in Quantum Physics**

Alan Forrester

15 Linden Road

Northampton

NN3 2JJ

United Kingdom

alan_forrester@yahoo.co.uk

*Abstract*

Most attempts to argue for the second law of thermodynamics fail because (1) they use the unviable frequency theory of probability and (2) they do not explain why the arrow of time seen in experiments is aligned with the thermodynamic arrow of time. I use the decision theoretic interpretation of quantum probability from the many worlds interpretation of quantum mechanics to solve the probability problem. I then derive a correlation between the knowledge arrow of time and the entropy arrow of time using physical constraints on knowledge creation imposed by Popper's evolutionary theory of knowledge and the many worlds interpretation.

1 Introduction
2 Previous Attempts to Prove the Second Law
3 The Knowledge Arrow of Time

**1 Introduction**

In this paper I shall argue that in the many worlds interpretation of quantum mechanics (henceforth the many worlds theory) (Everett [1957], Deutsch [1997, 2002]) the growth of entropy is correlated with the growth of knowledge (that is, useful or explanatory information). Since the psychological arrow of time is associated with the growth of knowledge this argument proves a version of the second law of thermodyamics – we will see entropy grow in all experiments if we account for all of the entropy created in the experiment.

In Section 2 I explain why previous attempts to prove the second law have failed. The arguments all assume the frequency interpretation of probability – probability is the relative frequency over an infinite set of trials. Since they all assume this interpretation of probability they do not explain what happens in any real experiment because all real experiments consist of a finite set of trials. These arguments also do not explain why any purported asymmetry should match the psychological arrow of time, which means they do not explain the results that we see in experiments.

Section 3 is about the arrow of time associated with the growth of knowledge. Knowledge has to be instantiated in some system that accurately reflects some of the causal and structural properties of some system in the past but the system simulated by the knowledge bearing system did not contain information about the knowledge bearing system. Karl Popper [1979] showed that knowledge grows as a result of evolutionary processes – processes that involve generating variations on existing knowledge and then selecting among those variations. As such knowledge must be instantiated in some system that can interact with the outside world without being changed because otherwise it cannot survive selection pressure even in principle and this places limits on the physical properties of knowledge bearing systems.

In Section 4 I argue that quantum physics places constraints on the sort of information that can be copied – it must be instantiated in a property of a physical system that can be represented by a sum of orthogonal projectors.

In Section 5 I explain that the decision theoretic theory of probability in the many worlds theory avoids the problems of the frequency theory of probability. According to the many worlds theory the universe we see around us is only part of a much larger structure called the multiverse which consists, in part, of many physical systems that act approximately like the universe as described by classical physics. Using restrictions that the many worlds theory imposes on the flow of information in the multiverse it can be shown that the expectation value $\langle \hat{A} \rangle$ of an observable $\hat{A}$ in the relative state $\rho$ is given by the Born rule

$$\langle \hat{A} \rangle = tr(\rho \hat{A}) \tag{1}$$

where $tr(\bullet)$ is the trace function.

In Section 6 I argue that the growth of knowledge is correlated with the growth of entropy when the restrictions on information copying imposed by quantum physics are taken into account.

**2 Previous Attempts to Prove the Second Law**

Most attempts to prove the second law of thermodynamics hang on arguments the ergodic theorem. This argument is well known so I will rehearse it only briefly to highlight the problems I consider relevant, for a critical survey of these arguments, see Earman and Redei [1996].

For any system of interest we typically cannot get access to all of the details of the system. For example, we can only measure the rough position of a cloud of gas, not the exact position of all of its constituent molecules. We also cannot measure the energies of all of the individual molecules only the average energy of some large number of molecules. The gross features of a system that we can measure are called the macrostate.

The fine-grained features that we cannot measure are called the microstate. For any given macrostate $M_A$ there are many microstates $m_{Aa}$ that would produce the features of that macrostate where $A$ and $a$ are indices that label the states.

Then it is argued that there is some measure $\mu_{Aa}$ over the set of microstates that gives the probability that if we see a macrostate $M_A$ the system has the microstate $m_{Aa}$. The probability is defined as the relative frequency of the microstate in question over an infinite amount of time or an infinite number of trials. In classical physics a suitable system will evolve so that the current microstate $m_{Aa}$ determines the future microstate $m_{f_t(Aa)}$ where $f_t(\bullet)$ is some time dependent function over the set of indices. The argument then runs along the lines that a given system will probably evolve so that $\mu_{f_t(Aa)} > \mu_{Aa}$ and the states which have higher probability according to the measure also have higher entropy where the entropy is some function of the measure and so is also a function of the probability. And so entropy will increase with time, thus proving the second law, or so the theory goes.

There are at least two problems with this argument. First, we do not measure any system for an infinite amount of time or over an infinite number of trials. So predictions of the relative frequency over an infinite amount of time or an infinite number of trials are irrelevant to any real experiment. Second, even if this problem did not cripple the argument we would have no explanation as to why the direction in time in which the parameter $t$ increases is also the direction that we happen to see when we do experiments. I solve the first problem by using the decision theoretic interpretation of probability in the many worlds theory. I solve the second problem by looking at the growth of knowledge as a physical process and showing that this process increases entropy.

**3 The Knowledge Arrow of Time**

Any piece of knowledge has to be instantiated in some physical object, e.g. – a human brain, a hard disc on a computer or the genes of an animal. If one physical system $S_1$ contains knowledge about another system $S_2$ then $S_1$ must contain a (possibly incomplete) description of $S_2$. In addition since some physical process generated this knowledge there must be a system that contains information about the knowledge creation processes: $S_3$. So $S_1$ and $S_3$ contain information about $S_2$ but not vice versa. So it seems reasonable to say that there is an arrow of time connected to the growth of knowledge. Before knowledge is created about $S_2$ there are no systems other than $S_2$ that contain knowledge about $S_2$ and afterward there are systems other than $S_2$ that contain information about $S_2$. So the time at which $S_1$ and $S_3$ contain information about $S_2$ is later than the time at which they do not contain information about $S_2$ according to the knowledge arrow of time. For example, a bird's wing and its genes contain information about the laws of aerodynamics but before wings evolved the air did not contain information about bird wings or genes. In this paper I argue that in the many worlds theory if there is a knowledge arrow of time of the kind described in this paragraph then the thermodynamic arrow of time is aligned with it.

Systems sometimes instantiate knowledge about things that are not physical objects, like prime numbers, so is it not true that some knowledge cannot be interpreted as information about a physical system instantiated in some other physical system? I shall argue that this can be reconciled with the knowledge arrow of time using mathematical knowledge as an example. Mathematical knowledge must be created by proposing that the properties of some physical object, like a mathematician's brain or a computer programme, are similar in some respects to the properties of the mathematical objects invoked in the proof. So a given piece of mathematical knowledge can be interpreted in two ways. First, it can be interpreted as knowledge of some mathematical objects. Second, it can be interpreted as knowledge about the physical properties of the objects that supposedly mirror the properties of the mathematical objects we are interested in and so the knowledge arrow of time described above applies in this case too.

Charles H. Bennett [1994] invented a measure of the knowledge in a physical object in terms of the number of steps that a Turing machine would have to take to describe that object – logical depth. (Strictly speaking the logical depth is actually the harmonic mean of the number of steps required by all of the different types of Turing machines.) Deutsch [1985] and Neilsen [2006] generalised the classical logical depth for the quantum theory of computation. In the quantum theory of computation a quantum computer can simulate any quantum system to any desired level of accuracy and so in quantum physics all systems can be considered as being the output of a quantum computation and so the idea of quantum logical depth can be applied to all quantum systems, which means that if the many worlds theory is correct all physical systems have some quantum logical depth. So in the many worlds theory knowledge in the sense I have described in this Section is an attribute of physical objects and this provides a link between the knowledge and physics. Deutsch [1985] suggested proving the second law by showing that increasing quantum logical depth is correlated with the growth of entropy. I follow this suggestion by the indirect route of showing that the evolutionary processes that are required to generate knowledge increase entropy.

Karl Popper [1979, Chapters 1 and 7] has shown that only processes that create variations on current knowledge and then select among those variations (evolutionary processes), can create knowledge. Human beings create knowledge by noticing problems with their current theories or habits, proposing changes to that knowledge to fix those problems (variation) and then criticising those proposals in the light of other knowledge (selection), which can include knowledge of results of observations. In biological evolution a mutation in an organism's germ line cells can change the phenotypes of its descendants (variation). Some of those new phenotypes can make the organism carrying it better at propagating its genes than its competitors and so the genotypes that lead to those phenotypes become more common than their competitors (selection). Just to be clear, although Popper's epistemology is often thought of as applying only to the growth of scientific knowledge in fact it applies much more widely because the logic of the problem of how to gain information about the world that can serve a specific purpose does not change when we are not dealing with people who are deliberately trying to create new explanations, as Popper himself realised (Popper [1979, p. 261]):

> All this may be expressed by saying that the growth of our knowledge is the result of a process closely resembling what Darwin called 'natural selection'; that is, *the natural selection of hypotheses*: our knowledge consists, at every moment, of those hypotheses which have shown their comparative fitness by surviving so far in their struggle for existence; a competitive struggle which eliminates those hypotheses which are unfit.
>
> This interpretation may be applied to animal knowledge, pre-scientific knowledge, and to scientific knowledge….
>
> This statement of the situation is meant to describe how knowledge really grows. It is not meant metaphorically… From the amoeba to Einstein, the growth of knowledge is always the same: we try to solve our problems, and to obtain, by a process of elimination, something approaching adequacy in our tentative solutions.

According to Popper knowledge must be subjected to selection if it is going to grow and improve. This means it must be possible for knowledge to survive tests, which means it must be possible for a knowledge bearing system to remain unchanged after an interaction. So it must be possible to copy the information in a knowledge bearing system without destroying it. This argument links the growth of knowledge to the physical process of copying information and so it links the knowledge arrow of time to information copying physical processes. I shall argue that it also links the second law of thermodynamics to the growth of knowledge.

## 4 The Quantum Physics of Copying

In this section I shall give an argument showing that in the Heisenberg picture the only physical features of a quantum system that can be copied are features associated with sums of orthogonal sets of projectors. This argument is similar to restrictions derived by Zurek [2007] showing that only information represented by orthogonal vectors in Hilbert space can be copied due to the linearity of quantum mechanics. It is also related to the no-cloning theorem (Wootters and Zurek, [1982]) which showed that no quantum mechanical process can copy an arbitrary quantum state due to linearity – if a physical process copies eigenstates of an observable $\hat{A}$ it will not copy eigenstates of an observable $\hat{B}$ that does not commute with $\hat{A}$. In the Heisenberg picture a quantum system is described by a stationary state $\rho$ and evolving observables, which are both represented by Hermitian operators $\hat{A}(t)$. The observables evolve unitarily:

$$\hat{A}(t_2) = U^\dagger_{t_2,t_1} \hat{A}(t_2) U_{t_2,t_1}, \tag{2}$$

where

$$U^\dagger_{t_2,t_1} U_{t_2,t_1} = \hat{1} \tag{3}$$

and $\hat{1}$ is the unit observable.

Instead of observables we can use any operator that spans the vector space of the observables may describe a system and to avoid pre-judging the issue of what information can be copied I will use a different set of operators. Consider two systems $S_1$ and $S_2$ represented by operators on Hilbert spaces $H_1$ and $H_2$ respectively. $H_1$ and $H_2$ can be spanned by operators $S_{ab1}(t)$ and $S_{ef2}(t)$ with the algebra

$$\begin{aligned}
&S_{Aab1}(t)S_{Acd1}(t) = \delta_{bc}S_{Aad1}(t) \\
&S_{Aef2}(t)S_{Agh2}(t) = \delta_{fg}S_{Aeh2}(t) \\
&\left[S_{Aab1}(t), S_{Aef2}(t)\right] = 0 \\
&\sum_a S_{Aaa1}(t) = \hat{1} \\
&\sum_e S_{Aee2}(t) = \hat{1}
\end{aligned} \qquad (4)$$

where the indices on the $S_{Aab1}(t)$ are integers in $1K\ N_1$ and the indices on the $S_{Aef2}(t)$ are integers in $1K\ N_2$. The $S_{Aaa1}(t)$ are orthogonal projectors $\hat{P}_{Aa1}(t)$ and so they are the projectors of some observable $\hat{A}_1(t) = \sum_a \alpha_{a1}\hat{P}_{Aa1}(t)$ of $S_1$ and the $S_{Aee2}(t)$ are orthogonal projectors $\hat{P}_{Ae2}(t)$ and so they are the projectors of some observable $\hat{A}_2(t) = \sum_e \alpha_{e2}\hat{P}_{Ae2}(t)$ of $S_2$. The $S_{Aab1}(t)$ can be written as linear combinations of observables with complex coefficients so they too change unitarily.

Any operator $B_1(t)$ describing some physical characteristic $S_1$ can be written as

$$B_1(t) = \sum_{ab} \beta_{ab} S_{Aab1}(t), \qquad (5)$$

where the $\beta_{ab}$ are complex c-numbers. $B_1(t)$ changes unitarily just as the $S_{Aab1}(t)$ do. If $B_1(t)$ is copied the copying process leaves $B_1(t)$ the same. So if the copying takes place between $t_1$ and $t_2$ there is some unitary operator $U_{t_2,t_1}$ such that

$$B_1(t_2) = \sum_{ab} \beta_{ab} S_{Aab1}(t_2)$$
$$= U^\dagger_{t_2,t_1} B_1(t_1) U_{t_2,t_1} = B_1(t_1) = \sum_{ab} \beta_{ab} S_{Aab1}(t_1) \quad (6)$$

So if the $\beta_{ab}$ are different from one another I will say that $B_1(t)$ is non-degenerate. The copying process leaves a non-degenerate $B_1(t)$ the same if

$$U^\dagger_{t_2,t_1} S_{Aab1}(t_1) U_{t_2,t_1} = S_{Aab1}(t_1). \quad (7)$$

If the $\beta_{ab}$ are not all distinct I will say that $B_1(t)$ is degenerate and the copying process leaves $B_1(t)$ the same if

$$U^\dagger_{t_2,t_1} S_{Aab1}(t_1) U_{t_2,t_1} = S_{Acd1}(t_1) \quad (8)$$

where $\beta_{ab} = \beta_{cd}$.

I will consider the case where $B_1(t)$ is non-degenerate first. Any unitary operator $U_{t_2,t_1}$ may be written as

$$U_{t_2,t_1} = \sum_{cdef} \gamma_{cdef} S_{Ace1}(t_1) S_{Adf2}(t_1) \quad (9)$$

where it can be shown using (3) that the $\gamma_{cdef}$ are complex numbers such that

$$\sum_{cd} \gamma_{abcd} \gamma^*_{efcd} = \delta_{ae} \delta_{bf}. \quad (10)$$

(4), (7) and (9) give

$$U^\dagger_{t_2,t_1} S_{Aab1}(t_1) U_{t_2,t_1} = \sum_{cdfgh} \gamma_{cdaf} \gamma^*_{ghbf} S_{Acg1}(t_1) S_{Adh2}(t_1) = S_{Aab1}(t_1). \quad (11)$$

From (11) $c = a$ and $g = b$ which implies that

$$U_{t_2,t_1} = \sum_{cd} \hat{P}_{Ac1}(t_1) u_{c2}, \qquad (12)$$

where the $u_{c2}$ are unitary operators acting on $\mathsf{H}_2$. (12) gives

$$U^\dagger_{t_2,t_1} S_{Aab1}(t_1) U_{t_2,t_1} = S_{Aab1}(t_1) u^\dagger_{a2} u_{b2}, \qquad (13)$$

so either $u_{a2} = u_{b2}$ or $a = b$. In the former case $U_{t_2,t_1}$ factorises and so does not describe an interaction between the two systems and so obviously does not represent copying. In the latter case $B_1(t)$ is a linear combination of members of an orthogonal set of projectors with complex coefficients.

What if $B_1(t)$ is degenerate? In this case

$$U_{t_2,t_1} = \sum_{cd} \hat{P}_{Ac1}(t_1) u_{c2} u_{\pi 1}, \qquad (14)$$

where

$$u_{\pi 1} = \prod_e \sum_{d_e \in D_e} S_{A\pi_e(d_e)d_e}(t_1), \qquad (15)$$

$D_1, D_2 \mathsf{K}$ are sets of indices that can be swapped without changing $B_1(t)$ and $\pi_1, \pi_2 \mathsf{K}$ are permutation functions on $D_1, D_2 \mathsf{K}$ respectively. So degeneracy does not affect the argument that $B_1(t)$ is a linear combination of an orthogonal set of projectors with complex coefficients. Any information that can be copied must be instantiated in a physical quantity represented by an operator that is a linear combination of orthogonal projectors with complex coefficients. This is a non-trivial restriction, which some physical attributes of a system will not satisfy. For example, all quantities represented by an operator $S_{Aab1}(t)$ with $a \neq b$ does not satisfy this restriction. Observables are usually assumed to be Hermitian operators, that is linear combinations of orthogonal projectors with real coefficients. Since this will not affect the substance of the argument I will give I will use Hermitian operators to represent information carrying physical quantities for the rest of this paper.

**5 Decision Theoretic Probability in the Many Worlds Theory**

If the many worlds theory is true then it is often the case that a measurement will have more than one outcome. Deutsch [1999], Wallace [2003], Zurek [2005] and Forrester [2007] have given arguments that can be interpreted as saying that the many worlds theory can incorporate probability as a weight that is attached to each branch by a rational decision theoretic agent – an agent who assigns 'values' to different games in a way that satisfies the 'rationality' requirement, namely that for any two games $G_1$ and $G_2$ either the value of $G_1$ is greater than that of $G_2$, or it is less than that of $G_2$ or it is equal in value to $G_2$. If the value of $G_1$ is greater than that of $G_2$, the agent will willingly give up an opportunity to experience $G_2$ for an opportunity to experience $G_1$ and if the ranking is reversed his preferences will reverse too. If their value is equal he will be indifferent between experiencing $G_1$ and experiencing $G_2$. It can be shown that the only set of values that obeys the rationality rules is the Born rule. I will explain how this argument works leaving out some of the details following Forrester [2007].

First I will outline the sense in which the multiverse can be said to contain many universes and then I will explain measurement theory in the many worlds theory. In the many worlds theory a universe is a structure within the multiverse that resembles the universe of classical physics. A system $S_1$ has observables

$$\hat{A}_1(t) = \sum_a \alpha_{a1} \hat{P}_{Aa1}(t)$$
$$\hat{C}_1(t) = \sum_c \chi_{c1} \hat{P}_{Cc1}(t) \qquad (16)$$

where

$$\hat{P}_{Cc1}(t) = \sum_{de} \eta_{cde1} S_{Ade1}(t) \qquad (17)$$

and the $\eta_{cde1}$ are complex c-numbers such that $\hat{P}_{Cc1}(t)\hat{P}_{Cb1}(t) = \delta_{bc}\hat{P}_{Cc1}(t)$.

Suppose that between times $t_1$ and $t_2$ $S_1$ evolves so that

$$U_{t_1,t_2} = \sum_b S_{A\pi(b)b}(t_1), \qquad (18)$$

where $\pi(\bullet)$ is a function that permutes the integers 1K $N_1$ then

$$\hat{A}_1(t_2) = \sum_a \alpha_{a1} \hat{P}_{A\pi(a)1}(t_1)$$
$$\hat{C}_1(t_2) = \sum_c \chi_{c1} \eta_{cde} S_{A\pi(d)\pi(e)}(t_1) \qquad (19)$$

This evolution permutes the projectors of $\hat{A}_1(t)$. A Turing machine that obeys classical physics, as defined by Turing [1936], permutes integers and so this evolution performs $N_1$ classical computations (computations that could be performed on a classical Turing machine) on $\hat{A}_1(t)$. By contrast, this evolution does not permute the projectors of an arbitrary observable $\hat{C}_1(t)$ of $S_1$ as illustrated by equation (19). So I will say that the computation on $\hat{A}_1(t)$ contains $N_1$ branches of the multiverse – that is, $N_1$ structures that look classical to any observers within the branch. The evolution of $\hat{C}_1(t)$ does not constitute $N_1$ branches. If a branch persists over a large region of space and time then I will say that it is a universe. There is no guarantee that the branch will persist because in principle the system might start evolving such that the evolution operator is not of the form (18).

I shall now briefly discuss measurement. A second system $S_2$ has observables

$$\hat{A}_2(t) = \sum_a \alpha_{a2} \hat{P}_{Aa2}(t)$$
$$\hat{C}_2(t) = \sum_c \chi_{c2} \hat{P}_{Cc2}(t) \quad (20)$$
$$\hat{P}_{Cc2}(t) = \sum_{de} \eta_{cde2} S_{Ade2}(t)$$

where the c-numbers $\eta_{cde2}$ are defined similarly to the c-numbers the $\eta_{cde1}$. During a perfect measurement $S_1$ and $S_2$ evolve so that

$$U_{t_2,t_1} = \sum_{a,b} \hat{P}_{Aa1}(t_1) S_{Aba \oplus_{N_1} b}(t_1) \quad (21)$$

where $a \oplus_{N_1} b = (a+b) \mod N_1$ and this gives

$$\hat{A}_1(t_2) = \sum_a \alpha_{a1} \hat{P}_{Aa1}(t_1)$$
$$\hat{A}_2(t_2) = \sum_{a,b} \alpha_{a2} \hat{P}_{Aa1}(t_1) \hat{P}_{Aa \oplus_{N_1} b 2}(t_1)$$
$$\hat{C}_1(t_2) = \sum_{abcd} \chi_{c1} \eta_{1cad} S_{Aad1}(t_1) S_{Aa \oplus_{N_1} bd \oplus_{N_1} b 2}(t_1) \quad (22)$$
$$\hat{C}_2(t) = \sum_c \chi_{c2} \eta_{1cbd} \hat{P}_{Aa1}(t_1) S_{Aa \oplus_{N_1} bd \oplus_{N_1} b 2}(t_1)$$

So the perfect measurement perfectly correlates $\hat{A}_1(t)$ with $\hat{A}_2(t)$ but does not perfectly correlate $\hat{A}_1(t)$ with $\hat{C}_2(t)$ or $\hat{C}_1(t)$ with $\hat{C}_2(t)$. So the measurement picks out $\hat{A}_1(t)$ and $\hat{A}_2(t)$ as preferred observables in which correlations are created between $S_1$ and $S_2$.

In the many worlds theory a single branch can evolve each branch at the start of the evolution is associated with many branches at the end of the evolution. For example, if $S_1$ and $S_2$ evolve as described in equations (21) and (22) above between $t_1$ and $t_2$ and then $\hat{A}_1(t)$ evolves so that $\hat{A}_1(t_3) = \hat{C}_1(t_2)$ and $\hat{A}_1(t)$ is perfectly measured onto $\hat{A}_2(t)$ between $t_3$ and $t_4$ then

$$U_{t_4,t_3} = \sum_{a,b} \hat{P}_{Aa1}(t_3) S_{Aba \oplus_{N_1} b2}(t_3)$$
$$\hat{A}_2(t_4) = \sum_{ab} \alpha_{2b} \hat{P}_{Aa1}(t_3) \hat{P}_{Aa \oplus_{N_1} b2}(t_3) \qquad (23)$$
$$= \sum_{abcde} \alpha_{2b} \eta_{aea \oplus_{N_1} b2} S_{Aea \oplus_{N_1} b1}(t_1) S_{Ae \oplus_{N_1} da \oplus_{N_1} b \oplus_{N_1} c2}(t_1)$$

The measurement still produces a branching structure involving $\hat{A}_1(t)$ and $\hat{A}_2(t)$. However, any particular branch at $t_1$ is not associated with a single branch at $t_4$ but with multiple branches. As such not all of the classical information that was recorded in $\hat{A}_2(t)$ before the measurement has been recovered from $S_2$ and so it has not acted as a stable channel for classical information. The fact that many quantum systems do act as stable channels for classical information is explained by the theory of decoherence, see Ollivier et al. [2004, 2005] and Zurek [1991].

The evolution of the observables by itself is not enough to make predictions about the results of experiments. An observer can experience only one of the universes and so to make predictions he must know what universe he is in and what observables are sharp in that universe – this information is contained in the relative Heisenberg state $\rho$ a Hermitian operator with unit trace. All of the information that an observer can access about an observable $\hat{A}(t)$ is contained in the product $\rho\hat{A}(t)$.

Using the restrictions that the measuring process places on what information can be copied from one system to another it can be shown that a quantum game consists of (1) an observable $\hat{A}(t)$, (2) a relative state $\rho$ and (3) a payoff function $P$ that maps each of the eigenvalues of $\hat{A}(t)$ to the payoff the agent gets from experiencing that eigenvalue and that the game whose payoff function does not change the eigenvalues assigns expectation values to observables according to the Born rule (1). This argument provides a non-frequency theoretic theory of probability.

**6 Knowledge and Entropy in Quantum Mechanics**

*6.1 The state of a knowledge bearing system*

The Schrödinger picture gives the same expectation values for observables as the Heisenberg picture and makes calculations somewhat easier although it is worse at tracking the flow of information (Deutsch and Hayden [2000]). I have now derived the constraints on the flow of information that I will need for my argument and so for ease of calculation from now on I will use the Schrödinger picture. The evolving global Schrödinger state $\rho(t)$ starts out equal to the static Heisenberg state at $t = 0$ and then evolves so that

$$\rho(t_2) = U_{t_2,t_1} \rho(t_1) U^{\dagger}_{t_2,t_1}. \tag{24}$$

In the Schrödinger picture the observables are static and are equal to the corresponding evolving Heisenberg observables at $t = 0$. At a given time $t_0$ the expectation values of measurements on a subsystem $S$ can be calculated by taking the reduced state $\rho_S(t_0)$ of $S$ given by

$$\rho_S(t_0) = tr_{\text{not } S}(\rho(t_0)) \tag{25}$$

where $tr_{\text{not } S}(\bullet)$ is the partial trace over all of the Hilbert spaces of systems other than $S$. In general the reduced state evolves so that

$$\rho_S(t_2) = tr_{\text{not } S}(U_{t_2,t_1} \rho(t_1) U^{\dagger}_{t_2,t_1}). \tag{26}$$

In Section 3 I argued that knowledge has to be information that can be copied. In Section 4 I showed that information that can be copied is instantiated in physical quantities that are represented by Hermitian operators – linear combinations of orthogonal projectors. Now consider a quantum system $S$ with two subsystems $S_1$ and $S_2$ in the state

$$\rho(t_1) = \sum_{ab} p_{ab} |\phi_{ab}\rangle_{12\ 12}\langle\phi_{ab}| \tag{27}$$

where

$$|\phi_{ab}\rangle_{12\ 12}\langle\phi_{ab}| = \sum_{cdef} \lambda_{abcd} \lambda_{abef} |c\rangle_{1\ 1}\langle e||d\rangle_{2\ 2}\langle f|, \tag{28}$$

with $|c\rangle_{1\ 1}\langle e||g\rangle_{1\ 1}\langle h| = \delta_{eg}|c\rangle_{1\ 1}\langle h|$, $|d\rangle_{2\ 2}\langle f||g\rangle_{2\ 2}\langle h| = \delta_{fg}|d\rangle_{2\ 2}\langle h|$ and the $\lambda_{abcd}$ are real numbers such that $\sum_{cd}\lambda^2_{abcd} = 1$. The $\lambda_{abcd}$ can be assumed to be real because of the Schmidt decomposition. The reduced states of $S_1$ and $S_2$ are

$$\rho_1(t_1) = \sum_{abcde} p_{ab} \lambda_{abcd} \lambda_{abed} |c\rangle_{1\ 1}\langle e|$$
$$\rho_2(t_1) = \sum_{abcdf} p_{ab} \lambda_{abcd} \lambda_{abcf} |d\rangle_{2\ 2}\langle f| \quad . \tag{29}$$

If the observable with projectors $|\phi_{ab}\rangle_{12\ 12}\langle\phi_{ab}|$ is measured then the state of $S$ will remain unchanged. But if an observable with the same orthogonal projectors as $\rho_1(t_1)$ is measured or an observable with the same orthogonal projectors as $\rho_2(t_1)$ is measured or they are both measured then the state of $S$ changes unless $|\phi_{ab}\rangle_{12\ 12}\langle\phi_{ab}|$ is a product of those projectors. So if $|\phi_{ab}\rangle_{12\ 12}\langle\phi_{ab}|$ is not a product of orthogonal projectors on the Hilbert spaces of $S_1$ and $S_2$ then $S$ cannot be treated as a knowledge bearing system along with $S_1$ and $S_2$ since $S$ changes under measurements of $S_1$ and $S_2$. So any knowledge bearing system that has knowledge bearing subsystems must have a state of the form

$$\rho(t_1) = \sum_{ab} p_{ab} |a\rangle_{1\ 1}\langle a| |b\rangle_{2\ 2}\langle b| . \tag{30}$$

It is important to note that not all states of a bipartite system are of the form (30). For example, the two-qubit state

$$\sigma = \tfrac{1}{2}(|00\rangle + |11\rangle)(\langle 00| + \langle 11|) \tag{31}$$

cannot be written in the form (30), i.e. – it cannot be written as a sum of unentangled pure states of the two qubits. A similar argument holds for knowledge bearing systems with more than two knowledge bearing subsystems. The state for a knowledge bearing system with $n$ subsystems would be

$$\rho(t_1) = \sum_{a_1 \mathrm{K} a_n} p_{a_1 \mathrm{K} a_n} \prod_j |a_j\rangle_{j\ j}\langle a_j| . \tag{32}$$

Since knowledge arises from selecting among different variations on a piece of knowledge, a knowledge bearing system $S$ will often contain knowledge bearing subsystems where $S$ contains information about the variations being compared and the results of the comparison.

*6.2 Entropy and decision theoretic probabilities*

I shall now show that the entropy of a system $S$ is the difference between the maximal information carrying capacity of a physical system and its actual information carrying capacity if the only systems we are allowed to use other than $S$ are in pure states and are not entangled with $S$. This provides an interpretation of entropy in terms of decision theoretic probabilities. Suppose that there are three systems $S_1, S_2$ and $S_3$ all described by operators on a Hilbert space of dimension $N$. I will show that if some system $S_1$ has the state $|a\rangle_1{}_1\langle a|$ for some number $a$ in the sequence 1K $N$ then if system $S_2$ is in the reduced Schrödinger state $\rho_2(t)$ and the agent using $S_2$ as a channel knows that it is in this state, if information is transferred first to $S_2$ and then to a system $S_3$, which is in a blank state $|0\rangle_3{}_3\langle 0|$, the maximum amount of information that can be transferred is

$$I_{max}(\rho_2(t)) = \log N - S(\rho_2(t)), \tag{33}$$

where the logarithms are taken to base 2 here and throughout the paper and

$$S(\rho_2(t)) = -\rho_2(t)\log \rho_2(t). \tag{34}$$

At time $t_1$ the state $\rho_2(t_1)$ may be written as

$$\rho_2(t_1) = \sum_b p_b |b\rangle_{2\,2}\langle b|. \tag{35}$$

The state of $S_1, S_2$ and $S_3$ at $t_1$ is

$$\rho(t_1) = |a\rangle_{1\,1}\langle a| \sum_b p_b |b\rangle_{2\,2}\langle b| |0\rangle_{3\,3}\langle 0| \tag{36}$$

where the $|b\rangle_{2\,2}\langle b|$ are orthogonal. To use $S_2$ as a channel for copying information about $a$ the best that can be done is to do a perfect measurement of the observable with projectors $|c\rangle_{1\,1}\langle c|$ and then a perfect measurement of the observable with projectors $|b\rangle_{2\,2}\langle b|$ onto the observable with projectors $|d\rangle_{3\,3}\langle d|$. When this sequence of measurements is complete at time $t_2$ the state is

$$\begin{aligned}\rho(t_2) &= |a\rangle_{1\,1}\langle a| \sum_b p_b |\pi_a(b)\rangle_{2\,2}\langle \pi_a(b)| |\pi_a(b)\rangle_{3\,3}\langle \pi_a(b)| \\ &= |a\rangle_{1\,1}\langle a| \sum_b p_{\pi_a(b)} |b\rangle_{2\,2}\langle b| |b\rangle_{3\,3}\langle b| \end{aligned} \tag{37}$$

where the functions $\pi_a(\bullet)$ are permutations of $1\mathrm{K}\,N$ a different function for each possible value of $a$. So the reduced state of $S_3$ is

$$\rho_3(t_2) = \sum_b p_{\pi_a(b)} |b\rangle_{3\,3}\langle b|. \tag{38}$$

The maximal amount of information an observer could have about $S_2$ before it was used as a channel is the expectation values of its observables. And the maximal amount of information that can be read out is gained by permuting the probabilities so the maximal amount of information that $S_2$ can carry is a function of the numbers $p_b$.

Suppose that $S_2$ has two subsystems $S_{21}$ and $S_{22}$ such that

$$\rho_2(t_1) = \sum_{bc} p_{b21} p_{c22} |b\rangle_{21\,21}\langle b| |c\rangle_{22\,22}\langle c| \tag{39}$$

Then observing the result $b$ in a measurement of $S_{21}$ does not yield any information about what the result would be of a measurement on $S_{22}$ so the information $I$ gained from getting both results is the sum of the information given by each result separately:

$$I(p_{b21}p_{c22}) = I(p_{b21}) + I(p_{c22}). \tag{40}$$

There should not be large changes in the amount of information we get for small changes in the $p_b$ so as a function of $p_b$, $I$ must be continuous:

$$I(p_b) \propto \log p_b. \tag{41}$$

The probability of getting the result $b$ and the associated information is $p_b$ so

$$I_{\max}(\rho_2(t)) \propto \sum_b p_b \log p_b. \tag{42}$$

When

$$\rho_2(t_1) = |b_0\rangle_{2\ 2}\langle b_0| \tag{43}$$

where $b_0$ is some integer in the range 1K $N$, $S_2$ can hold as many bits as can be held in an integer in the range 1K $N$, i.e. $-\log N$ bits. When $S_2$ has the state

$$\rho_2(t_1) = \sum_b \tfrac{1}{N}|b\rangle_{2\ 2}\langle b| \tag{44}$$

permuting the numbers $p_b$ does not change the state and so no information can be conveyed in $S_2$ alone. $S_2$ could be part of a larger system carrying information but discussing this would bring in resources other than $S_2$. (42), (43) and (44) give

$$\begin{aligned} I_{\max}(\rho_2(t)) &= \log N + \sum_b p_b \log p_b \\ &= \log N - S(\rho_2(t)) \end{aligned} \tag{45}$$

There has been a controversy over whether or not the von Neumann entropy corresponds to the thermodynamic entropy based on analyses of a thought experiment by von Neumann [1932] (see Shenker [1999], Henderson [2003] and Shenker and Hemmo, [2007]). In the Appendix to this paper I show that these arguments against the von Neumann entropy do not work.

*6.3 Evolutionary processes and the growth of entropy*

Now I shall show that evolutionary processes lead to the growth of entropy. The growth of knowledge starts with two or more variations on the same knowledge and the results of previous comparisons between then and so starts with a knowledge bearing system $S$ with knowledge bearing subsystems. I shall show that selection between the different variations in the subsystems of $S$ increases the entropy of those subsystems. What if there are many different pieces of knowledge undergoing testing? A system containing many different pieces of knowledge has a state of the form (32) and any such state can be rewritten as a state of the form (30) so I can use a state of the form (30) without any significant loss of generality. So at the start of the selection process at $t_1$ the state of $S$ is

$$\rho(t_1) = \sum_{ab} p_{ab} |a\rangle_1{}_1\langle a| |b\rangle_2{}_2\langle b|. \tag{46}$$

The reduced states of the subsystems at $t_1$ are

$$\begin{aligned}\rho_1(t_1) &= \sum_{ab} p_{ab} |a\rangle_1{}_1\langle a| \\ \rho_2(t_1) &= \sum_{ab} p_{ab} |b\rangle_2{}_2\langle b|\end{aligned} \tag{47}$$

The selection process is a computation involving both $S_1$ and $S_2$. This computation must be arranged using current knowledge, which is bound to be imperfect. So even in the case of a classical, i.e. – decoherent, computation it will be impossible for epistemological reasons to arrange for the computation to leave $S_1$ and $S_2$ in a state of the form (46). A slightly more specific argument might help motivate this idea. Knowledge of the orthogonal projectors of the density matrix must be created by a process that produces conjectures about the state and then performing tests of those conjectures. However, those tests do not give perfect access to the expectation values of observables. If the same observables were measured every time in any particular universe they give access to the relative frequencies of different results. The larger the number of trials the more rational it will be (in the decision theoretic sense) to expect that the relative frequencies are close to the probabilities but the relative frequencies will not match the probabilities perfectly (see Forrester [2007, Section 4]). In reality the experiment is performed with imperfect knowledge from previous rounds of conjecture and criticism and so the tester's knowledge of what observable he measures in any particular test will be imperfect and this introduces another source of error. So if the selection process takes place between $t_1$ and $t_2$ the state at $t_2$ is of the form

$$\rho(t_2) = \sum_{ab} p_{ab} |\phi_{ab}\rangle_{12}\, {}_{12}\langle \phi_{ab}| \qquad (48)$$

where

$$|\phi_{ab}\rangle_{12}\, {}_{12}\langle \phi_{ab}| = \sum_{cdef} \lambda_{abcd} \lambda_{abef} |c\rangle_1\, {}_1\langle e| |d\rangle_2\, {}_2\langle f|. \qquad (49)$$

The reduced states of the subsystems at $t_2$ are

$$\rho_1(t_2) = \sum_{abcde} p_{ab} \lambda_{abcd} \lambda_{abed} |c\rangle_{1\ 1}\langle e|$$
$$\rho_2(t_2) = \sum_{abcdf} p_{ab} \lambda_{abcd} \lambda_{abcf} |d\rangle_{2\ 2}\langle f|. \quad (50)$$

The expectation values of observables of $S_1$ are now correlated with those of $S_2$ and they contain information about one another that they did not contain before the interaction and that is what constitutes $t_2$ being later than $t_1$ in this theory. This argument assumes that the knowledge bearing systems evolve unitarily which seem too restrictive an assumption but the whole world evolves unitarily in the many worlds theory and so we can always expand one of the knowledge bearing systems to include the rest of the universe and the whole system $S$ will then evolve unitarily. Let

$$|v_c\rangle_{1\ 1}\langle v_c| = \sum_{de} \xi_{cd} \xi_{ce} |d\rangle_{1\ 1}\langle e| \quad (51)$$

where the $\xi_{cd}$ are real c-numbers such that $\sum_c \xi_{cd} \xi_{fd} = \delta_{cf}$, then the $|v_c\rangle_{1\ 1}\langle v_c|$ are an orthogonal set of projectors and

$$|v_c\rangle_{1\ 1}\langle v_c||a\rangle_{1\ 1}\langle a||v_c\rangle_{1\ 1}\langle v_c| = \sum_{de} \xi_{cd} \xi_{ce} \xi_{ca}^2 |d\rangle_{1\ 1}\langle e|. \quad (52)$$

Equations (51), (52) and (47) give

$$\sum_c |v_c\rangle_{1\ 1}\langle v_c|\rho_1(t_1)|v_c\rangle_{1\ 1}\langle v_c| = \sum_{abcde} p_{ab} \xi_{cd} \xi_{ce} |d\rangle_{1\ 1}\langle e|. \quad (53)$$

It is well known that (Nielsen and Chuang [2000, Theorem 11.9, p.515]):

$$S\left(\sum_c |v_c\rangle_{1\ 1}\langle v_c|\rho_1(t_1)|v_c\rangle_{1\ 1}\langle v_c|\right) \geq S(\rho_1(t_1)). \quad (54)$$

Equation (50) is of the same form as (53) and so

$$S(\rho_1(t_2)) \geq S(\rho_1(t_1)) \quad (55)$$

and by symmetry

$$S(\rho_2(t_2)) \geq S(\rho_2(t_1)). \tag{56}$$

After the computation has been performed the results have to be measured and this too will increase entropy by a similar argument. The only way to decrease the entropy increase in the knowledge bearing system would be for the interaction to be reversed, which would delete the acquired knowledge or to export the entropy to another system by performing a swap operation on another system with less entropy. But this would just mean that the entropy of some other system would increase and so it would not allow the evasion of the second law. So the growth of knowledge is correlated with the growth of entropy.

**7 Discussion**

The argument given for the second law of thermodynamics in this paper depends on the many worlds theory and Popper's evolutionary theory of knowledge. The evolutionary theory of knowledge is important because it is a non-anthropomorphic and non-subjective theory of knowledge. This allows knowledge creation to be thought of as a physical process and so allows us to study its physical effects. It also specifies what sort of physical processes can generate knowledge – process that involve the production of variations on existing knowledge followed by selection among the competing variations.

The many worlds theory is useful for understanding the growth of knowledge and entropy for many reasons. The many worlds theory implies that quantum mechanics is a universal physical theory, that is, it is in the many worlds theory quantum mechanics applies to the whole universe. In the quantum theory of computation a universal quantum computer can simulate an physical system so if quantum mechanics applies to the whole universe any motion can be thought of as a computation and the final state of that motion can be thought of as the results of a computation. The many worlds theory also allows us to understand probability by studying limitations on the acquisition of information and so places limits on the ways in which a decision theoretic agent can place bets in order to win – he has to bet according to the Born rule. Nor do quantum probabilities hang on the frequency interpretation although the decision theoretic argument in the many worlds theory does explain why relative frequencies are relevant to experimental tests.

Some might object that my argument does not take error correction into account but error correction only reduces errors so that they do not prevent quantum computation from being scalable it does not eliminate them, for a survey of error correction, see Nielsen and Chuang [2000, Chapter 10]. Another possible objection is that this argument only shows that the entropy of subsystems of the universe has to increase in order for the generation of knowledge to take place but the entropy of the whole multiverse remains constant. However, we cannot manipulate the whole multiverse and the argument I gave shows that the entropy of the systems we can manipulate increases when knowledge grows. So the second law as argued for in this paper is relevant to what we can do and the entropy of the whole multiverse is not.

**Appendix Von Neumann Entropy and Thermodynamic Entropy**

There has been a controversy surrounding a thought experiment proposed by von Neumann [1932] to show that the von Neumann entropy is the quantum mechanical thermodynamic entropy. In the traditional pre-quantum version of the experiment an ideal gas starts at $t = 0$ in the left hand side of a box walled off by a partition. The partition is removed and it is so light that removing it does not involve doing any work. The gas expands into the right hand side of the box doing work and generating entropy $N \log 2$ where units have been chosen so that $k = \ln 2$ where $k$ is Boltzmann's constant. The gas is then compressed quasi-statically back to its former volume which reduces the entropy back to its value at the start of the experiment. Shenker [1999] and Shenker and Hemmo [2007] have argued that the von Neumann entropy does not change in this way and so it cannot be the thermodynamic entropy.

The experiment starts at $t = 0$ with a box divided by a partition. In the left side of the box there are some particles and each particle acts as a carrier for a qubit. Since the thermodynamic entropy depends on the position of gas molecules I will make a case that the entropy of the gas position changes in the way one would expect the thermodynamc entropy to change. The $j$th qubit $Q_j$ has been prepared in the state $|0_z\rangle_j {}_j\langle 0_z|$ – the projector of the Schrödinger picture observable

$$\sigma_{jz} = |0_z\rangle_j {}_j\langle 0_z| - |1_z\rangle_j {}_j\langle 1_z| \tag{57}$$

with eigenvalue 1. Also involved in the experiment are two measuring devices. One of these, $M_{xqj}$, measures the observable:

$$\sigma_{jx} = |0_x\rangle_{j\,j}\langle 0_x| - |1_x\rangle_{j\,j}\langle 1_x| \tag{58}$$

of the $Q_j$ where

$$\begin{aligned}
|0_x\rangle_{j\,j}\langle 0_x| &= \tfrac{1}{2}\left(|0_z\rangle_{j\,j}\langle 0_z| + |0_z\rangle_{j\,j}\langle 1_z| + |1_z\rangle_{j\,j}\langle 0_z| + |1_z\rangle_{j\,j}\langle 1_z|\right) \\
|1_x\rangle_{j\,j}\langle 1_x| &= \tfrac{1}{2}\left(|0_z\rangle_{j\,j}\langle 0_z| - |0_z\rangle_{j\,j}\langle 1_z| - |1_z\rangle_{j\,j}\langle 0_z| + |1_z\rangle_{j\,j}\langle 1_z|\right) \\
|0_z\rangle_{j\,j}\langle 0_z| &= \tfrac{1}{2}\left(|0_x\rangle_{j\,j}\langle 0_x| + |0_x\rangle_{j\,j}\langle 1_x| + |1_x\rangle_{j\,j}\langle 0_x| + |1_x\rangle_{j\,j}\langle 1_x|\right) \\
|1_z\rangle_{j\,j}\langle 1_z| &= \tfrac{1}{2}\left(|0_x\rangle_{j\,j}\langle 0_x| - |0_x\rangle_{j\,j}\langle 1_x| - |1_x\rangle_{j\,j}\langle 0_x| + |1_x\rangle_{j\,j}\langle 1_x|\right)
\end{aligned} \tag{59}$$

The other measuring device $M_{scj}$ measures which side of the box the carrier of the $j$th qubit is on. The state of the whole system: the qubits, the position of the molecules and the measuring devices is

$$\begin{aligned}
\rho(0) &= |0_z\rangle_{j\,j}\langle 0_z| |L\rangle_{cj\,cj}\langle L| |0\rangle_{Mxqj\,Mxqj}\langle 0| |0\rangle_{Mcj\,Mcj}\langle 0| \\
&= \tfrac{1}{2}\left(|0_x\rangle_{j\,j}\langle 0_x| + |0_x\rangle_{j\,j}\langle 1_x| + |1_x\rangle_{j\,j}\langle 0_x| + |1_x\rangle_{j\,j}\langle 1_x|\right) |L\rangle_{cj\,cj}\langle L| |0\rangle_{Mxqj\,Mxqj}\langle 0| |0\rangle_{Mcj\,Mcj}\langle 0|
\end{aligned} \tag{60}$$

At this point the entropy of the carrier position is zero – the same as the thermodynamic entropy. Between $t = 0$ and $t = 1$, $\sigma_{xj}$ is perfectly measured by $M_{xqj}$ and the state changes to

$$\rho(1) = \tfrac{1}{2}\begin{pmatrix} |0_x\rangle_{j\,j}\langle 0_x| |0\rangle_{Mxqj\,Mxqj}\langle 0| + |0_x\rangle_{j\,j}\langle 1_x| |0\rangle_{Mxqj\,Mxqj}\langle 1| + \\ |1_x\rangle_{j\,j}\langle 0_x| |1\rangle_{Mxqj\,Mxqj}\langle 0| + |1_x\rangle_{j\,j}\langle 1_x| |1\rangle_{Mxqj\,Mxqj}\langle 1| \end{pmatrix} |L\rangle_{cj\,cj}\langle L| |0\rangle_{Mcj\,Mcj}\langle 0| . \tag{61}$$

The reduced states of $Q_j$ and $M_{xqj}$ are

$$\rho_{Qj}(1) = \tfrac{1}{2}\left(|0_x\rangle_j{}_j\langle 0_x| + |1_x\rangle_j{}_j\langle 1_x|\right)$$
$$\rho_{Mqjx}(1) = \tfrac{1}{2}\left(|0\rangle_{Mxqj}{}_{Mxqj}\langle 0| + |1\rangle_{Mxqj}{}_{Mxqj}\langle 1|\right)$$
(62)

Between $t = 1$ and $t = 2$ the $\sigma_{xj}$ observable is measured onto the carrier position and so the state is

$$\rho(2) = \tfrac{1}{2}\begin{pmatrix} |0_x\rangle_j{}_j\langle 0_x||0\rangle_{Mxqj}{}_{Mxqj}\langle 0||L\rangle_{cj}{}_{cj}\langle L| \\ +|0_x\rangle_j{}_j\langle 1_x||0\rangle_{Mxqj}{}_{Mxqj}\langle 1||L\rangle_{cj}{}_{cj}\langle R| \\ +|1_x\rangle_j{}_j\langle 0_x||1\rangle_{Mxqj}{}_{Mxqj}\langle 0||R\rangle_{cj}{}_{cj}\langle L| \\ +|1_x\rangle_j{}_j\langle 1_x||1\rangle_{Mxqj}{}_{Mxqj}\langle 1||R\rangle_{cj}{}_{cj}\langle R| \end{pmatrix}|0\rangle_{Mcj}{}_{Mcj}\langle 0|$$
(63)

The reduced states of $Q_j$ and $M_{xqj}$ remain unchanged and the reduced state of the carrier becomes

$$\rho_{cj}(2) = \tfrac{1}{2}\left(|L\rangle_{cj}{}_{cj}\langle L| + |R\rangle_{cj}{}_{cj}\langle R|\right).$$
(64)

So its von Neumann entropy increases to $\log 2$. The molecules do not interact and it is easy to show that their total von Neumann entropy is $N \log 2$. At this point there are molecules on both sides of the container and the von Neumann entropy of the carrier position is the same as the thermodynamic entropy.

To quasi-statically compress the gas it is necessary to measure what side of the container each molecule is on. Between $t = 2$ and $t = 3$ the position of each molecule is measured:

$$\rho(3) = \tfrac{1}{2}\begin{pmatrix} |0_x\rangle_j{}_j\langle 0_x||0\rangle_{Mxqj}{}_{Mxqj}\langle 0||L\rangle_{cj}{}_{cj}\langle L||L\rangle_{Mcj}{}_{Mcj}\langle L| \\ +|0_x\rangle_j{}_j\langle 1_x||0\rangle_{Mxqj}{}_{Mxqj}\langle 1||L\rangle_{cj}{}_{cj}\langle R||L\rangle_{Mcj}{}_{Mcj}\langle R| \\ +|1_x\rangle_j{}_j\langle 0_x||1\rangle_{Mxqj}{}_{Mxqj}\langle 0||R\rangle_{cj}{}_{cj}\langle L||R\rangle_{Mcj}{}_{Mcj}\langle L| \\ +|1_x\rangle_j{}_j\langle 1_x||1\rangle_{Mxqj}{}_{Mxqj}\langle 1||R\rangle_{cj}{}_{cj}\langle R||L\rangle_{Mcj}{}_{Mcj}\langle R| \end{pmatrix}.$$
(65)

Then a unitary operation can be performed between $t = 3$ and $t = 4$ such that

$$\rho(4)= \tfrac{1}{2}|0_z\rangle_j{}_j\langle 0_z| |L\rangle_{cj}{}_{cj}\langle L| \begin{pmatrix} |0\rangle_{Mxqj}{}_{Mxqj}\langle 0| |L\rangle_{Mcj}{}_{Mcj}\langle L| + |0\rangle_{Mxqj}{}_{Mxqj}\langle 1| |L\rangle_{Mcj}{}_{Mcj}\langle R| \\ +|1\rangle_{Mxqj}{}_{Mxqj}\langle 0| |R\rangle_{Mcj}{}_{Mcj}\langle L| + |1\rangle_{Mxqj}{}_{Mxqj}\langle 1| |L\rangle_{Mcj}{}_{Mcj}\langle R| \end{pmatrix}. \quad (66)$$

The gas returns to its initial state and so has zero entropy. Each measuring device has entropy $N\log 2$. So the von Neumann entropy of the position of the gas increases to $N\log 2$ as a result of expanding into the right hand side of the box and then the entropy falls back to its former value as it is compressed back into the left hand side of the container. So the entropy changes just as it would according to thermodynamics. So this specific argument against the von Neumann entropy as the thermodynamic entropy fails.